\documentclass[epj]{svjour}
\usepackage{graphics}
\usepackage{bm}   
\usepackage{color}
\usepackage{amssymb}
\usepackage{amsmath}
\usepackage{balance}
\usepackage{subfigure}

\begin{document}
	
	\title{Rogue waves on the double periodic background in Hirota equation}
	
	\author{N.~Sinthuja \inst{1}, K.~Manikandan \inst{1} \and M. Senthilvelan \inst{1,} \thanks{\emph{e-mail:velan@cnld.bdu.ac.in}}}
	\institute{Department of Nonlinear Dynamics, Bharathidasan University, Tiruchirappalli 620024, Tamil Nadu, India.}

	\abstract{
		We construct rogue wave solutions on the double periodic background for the Hirota equation through one fold Darboux transformation formula.  We consider two types of double periodic solutions as seed solutions.  We identify the squared eigenfunctions and eigenvalues that appear in the one fold Darboux transformation formula through an algebraic method with two eigenvalues.  We then construct the desired solution in two steps.  In the first step, we create double periodic waves as the background.  In the second step, we build rogue wave solution on the top of this double periodic solution. We present the localized structures for two different values of the arbitrary parameter each one with two different sets of eigenvalues.}
	\titlerunning{Rogue waves on the double-periodic background in Hirota equation}
	\authorrunning{N. Sinthuja et al.}
	\maketitle

\section{Inroduction}
The nonlinear Schr\"odinger (NLS) equation describes the wave phenomena in the ocean \cite{dudley,fujim,degas,resi} and the propagation of optical pulse in optical fiber \cite{ysong,copie,onorato,grimshaw,akmdiev}.  Considering the external factors such as depth of the sea, bottom friction and viscosity in the ocean and the femtosecond pulse propagation in fibers, several higher order NLS equations have been introduced in the literature \cite{schen,vithya,zlan,yyang,zhangx,crabb}.  One such higher-order NLS equation is the Hirota equation.  In dimensionless form, the Hirota equation reads \cite{crabb},
\begin{equation}
i \psi_t+\alpha_2(\psi_{xx}+2|\psi|^2\psi)-i \alpha_3(\psi_{xxx}+6|\psi|^2\psi_x)=0,
\label{HE1} 
\end{equation} 
where the function $\psi =\psi(x,t)$ denotes the complex wave envelope,  $\alpha_2$ and $\alpha_3$ are arbitrary parameters.  The subscripts $t$ and $x$ stand for partial derivatives with respect to temporal and spatial variables, respectively. When $\alpha_2=1,\alpha_3=0$, Eq. $(\ref{HE1})$ reduces to the standard NLS equation.  During the past two decades, several works have been devoted to construct and analyze the characteristics of localized solutions in the Hirota equation \cite{anki:soto,wang,chow:anki,mlix,jcen,chen6}.  Multisolitons, breathers and rogue wave (RW) solutions have been constructed for the Hirota equation using different techniques, see for example Refs. \cite{wang,chow:anki,lli,tao}.  In these studies, RW solutions have been constructed only on the constant wave background.  

In nature, RWs can arise from a variety of backgrounds, including  constant background, multi-soliton background and periodic wave background \cite{agaf,mu,agaf1,mani,mani2}. Recently, interest has been shown to construct RW solutions on the periodic background.  Initially, the occurrence of RWs on periodic background has been brought out only through numerical schemes.  For example, computations of RWs on the periodic background were performed in \cite{kedziora}, where the authors have computed the solutions of Zakharov-Shabat spectral problem only numerically.  Similarly, RWs on the double-periodic background were constructed in \cite{calini} but only via a numerical scheme. The resulting wave patterns exhibit periodicity in both space and time.  Differing from these, very recently, Chen and his collaborators have reported an algorithm to construct RW solutions on the periodic wave background for the focusing NLS equation \cite{chen2}.  The same authors have also derived RW solutions on the double-periodic background for the focusing NLS equation using an effective algebraic method \cite{chen4}.   Subsequently, Peng et al have studied the characteristics of RWs on the periodic background for the Hirota equation \cite{peng}. However, the authors have reported RWs only for a limiting case.  The studies on RWs on the periodic background can be found potential applications both in experiment and theory in their relative fields \cite{rand,kim}.

In this work, we construct RW solutions on the double-periodic background for the Hirota equation.  We utilize one fold Darboux transformation formula and the method of nonlinearization of Lax pairs to derive these solutions.  To begin, we assume a constraint between the potential and squared eigenfunctions of the Lax system for the Hirota equation.  Using this constraint we identify Hamiltonians associated with the spatial and temporal parts of the Lax pair.  We notice that the Hamiltonian derived from spatial part of the Lax pair of Hirota equation matches with the one constructed for the NLS equation.  This is because the spatial part of the spectral problem of NLS equation and the Hirota equation are one and the same.  However, the Hamiltonian associated with the time part of the Hirota equation differs from NLS counterpart.  Since the spatial part is same, we obtain the same sets of differential constraints which connect the potential with eigenvalues that are given in Ref. \cite{chen4}.  We construct periodic eigenfunctions by solving the differential constraints in terms of known double periodic solutions of the Hirota equation.  Substituting these periodic eigenfunctions and the identified set of eigenvalues in the one-fold Darboux transformation formula, we create the double-periodic wave background.  We then construct another linearly independent solution of the same Lax equations for the same set of eigenvalues which in turn creates the RW solution on top of the double-periodic wave  background.  Since the time evolution part in the Lax pair of Hirota equation differs from the NLS equation the second independent solution which we construct for the Hirota equation differs from the NLS equation.  We present surface plots of $|\hat{r}|$ for two different values of $\alpha_3$ each one with two different sets of eigenvalues.  We find that the amplitude of RWs increases in the order of eigenvalues, such as $\lambda_2 < \lambda_1 < \lambda_3$ in the $(x-t)$ plane.  When we vary the value of the arbitrary parameter $\alpha_3$, we notice that the amplitude of RWs remains the same whereas the double-periodic background waves are localized in different positions in the $(x-t)$ plane.   

We organize our presentation as follows.  In Sec. 2, to begin, we present the Lax pair and one-fold Darboux transformation formula for the Hirota equation and two types of double periodic solutions of the Hirota equation.  We also briefly recall the method of constructing RW solution on the double periodic background.  In Sec. 3, we apply the algebraic method with two eigenvalues to Hirota equation and construct the RWs on the desired background.  In Sec. 4, we summarize our work.
\section{Lax pair and Darboux transformation}
The complete integrability of the nonlinear system (\ref{HE1}) is guaranteed by the presence of Lax pair.  Once the Lax pair is known, one can efficiently exploit the Darboux transformation technique to construct various localized solutions \cite{chow:anki,mlix,jcen,chen6} for the concerned equation.   With the help of Darboux transformation method, we can construct a new solution $\psi(x,t)=\hat{r}(x,t)$ by providing a suitable seed solution $\psi(x,t)=r(x,t)$.  The solution $\psi(x,t)=r(x,t)$ to the Hirota Eq. (\ref{HE1}) comes out from the compatibility condition $(\varphi_{xt}=\varphi_{tx})$ of the pair of linear equations on $\varphi$. The Lax pair for the Hirota equation is given by \cite{tao,peng}
\begin{align}
\varphi_{x}=U(\lambda,r) \varphi,\qquad U(\lambda,r)=\begin{pmatrix} \lambda & r\\ -\bar{r} & -\lambda \end{pmatrix},
\label{HE4}
\end{align}
and
\begin{align}
 &\varphi_{t} = V(\lambda,r) \varphi,\nonumber\\ &V(\lambda,r)=\lambda^3 \begin{pmatrix} 4\alpha_3  & 0\\ 0 & -4\alpha_3 \end{pmatrix}+\lambda^2 \begin{pmatrix} 2 i\alpha_2  & 4\alpha_3 r \\ -4\alpha_3 \bar{r} & -2 i \alpha_2 \end{pmatrix} \nonumber \\
& +~\lambda\begin{pmatrix} 2 \alpha_3 |r|^2  & 2\alpha_3 r_x+2 i \alpha_2 r \\ 2\alpha_3 \bar{r}_x-2i\alpha_2 \bar{r} & -2  \alpha_3 |r|^2\end{pmatrix} \nonumber \\ & +~\begin{pmatrix} i\alpha_2 |r|^2-\alpha_3(r \bar{r}_x-\bar{r} r_x)  & i\alpha_2 r_x+ \alpha_3( r_{xx}+2|r|^2 r )\\ i\alpha_2 \bar{r}_x- \alpha_3( \bar{r}_{xx}+2|r|^2 \bar{r} ) & -i\alpha_2 |r|^2+\alpha_3(r \bar{r}_x-\bar{r} r_x)\end{pmatrix},  
\label{HE5}
\end{align}
where $r$ is the potential and $\bar{r} $ denotes the complex conjugate of $r$. One can easily verify that the compatibility condition $U_t-V_x+[U,V]=0$ gives rise to Eq. $(\ref{HE1})$.   

\begin{figure*}[!ht]
	\begin{center}
		\subfigure[]
		{
		\resizebox{0.4\textwidth}{!}{\includegraphics{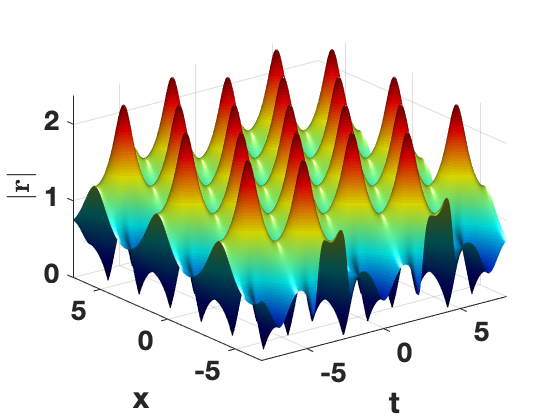}}
		\label{fig:1}
	}~~ 
\subfigure[]
{
		\resizebox{0.4\textwidth}{!}{\includegraphics{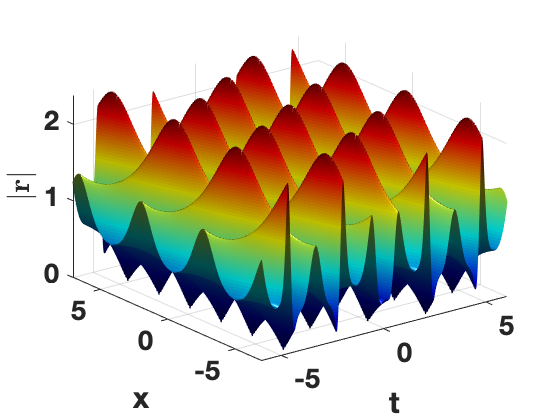}}
		\label{fig:2}
	}\\
	\subfigure[]
	{
		\resizebox{0.4\textwidth}{!}{\includegraphics{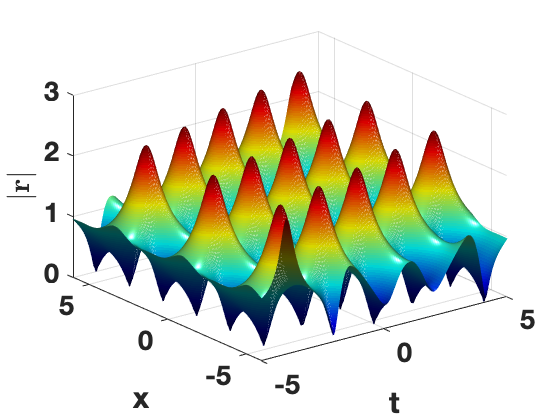}}
		\label{fig:3}
	}~~
	\subfigure[]
	{
		\resizebox{0.4\textwidth}{!}{\includegraphics{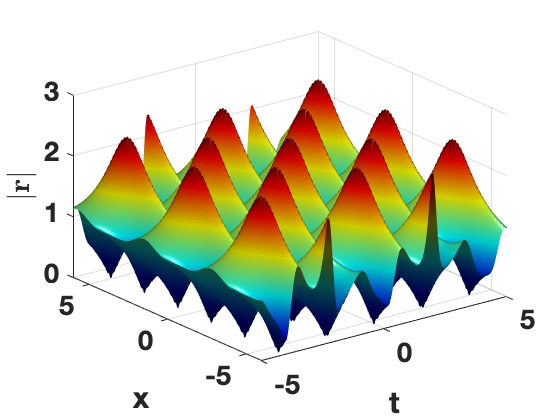}}
		\label{fig:4}
	}
	\end{center}
	\vspace{-0.3cm}
	\caption{Double periodic profile of (\ref{HE2}) for $k=0.9$ with (a) $\alpha_3=0.1667$ and (b) $\alpha_3=0.85$ and double periodic profile of (\ref{HE3}) for $k=0.9$ with (c) $\alpha_3=0.1667$ and (d) $\alpha_3=0.5$.}
	\label{dnnfig2}
\end{figure*}

The one-fold Darboux transformation of the Eq. (\ref{HE1}) reads \cite{chen2}
\begin{equation}
\hat{r}(x,t)=r(x,t)+\frac{2(\lambda_1+\bar{\lambda}_1)f_1 \bar{g}_1}{|f_1|^2+|g_1|^2},
\label{HE6}
\end{equation}
where $\varphi=(f_1(x,t), g_1(x,t))^T$ - a non-zero solution of the Lax pair $(\ref{HE4})$ and $(\ref{HE5})$ for a fixed eigenvalue $\lambda=\lambda_1$ and $r(x,t)$ and $\hat{r}(x,t)$ represents the seed and first iterated solutions of Eq. (\ref{HE1}), respectively.  With the chosen seed solution, the linear Eqs. $(\ref{HE4})$ and $(\ref{HE5})$ can be solved in a number of ways.  For example, it is trivial to integrate the Eqs. $(\ref{HE4})$ and $(\ref{HE5})$ with the plane wave background $(r(x,t)=e^{i(k x-\omega t)})$ and generate the first-order RW solution $(\hat{r}(x,t))$.  However, it is highly nontrivial to integrate the Eqs. $(\ref{HE4})$ and $(\ref{HE5})$ with double periodic solutions.  So one has to adopt a suitable methodology to integrate them with the genus-2 solutions as background.

Very recently double periodic solutions for the Hirota equation with free real parameters have been constructed in Ref. \cite{crabb}.  In our present study, we consider two types of double-periodic solutions which are given by the following rational functions of Jacobian elliptic functions sn, cn, and dn in \cite{crabb}:
\begin{subequations}
\label{HE2}
\begin{align} 
(i) \;\;\; r(x,t) = k \frac{\sqrt{1+k}\text{sn}(M t,k)-i C(x+v t)\text{cn}(M t,k)}{\sqrt{1+k}-C(x+v t)\text{dn}(M t,k)} e^{i b t},
\end{align}
\text{where the function $C(x+vt)$ is given by}
\begin{align}
C(x+vt) = \frac{\text{cn}(\sqrt{1+k}x,\kappa)}{\text{dn}(\sqrt{1+k}x,\kappa)}, \quad \kappa=\sqrt{\frac{1-k}{1+k}}.
\end{align}
\end{subequations}
\begin{subequations}
\label{HE3}
\begin{align}
(ii) \;\;\; r(x,t)=\frac{k\text{sn}(M t/k,k)-i S(x+v t)\text{dn}(M t/k, k)}{k(1-S(x+v t)\text{cn}(M t/k,k))}e^{ibt}, 
\end{align}
\text{where the function $S(x+vt)$ is given by}
\begin{align}
S(x+vt)=\sqrt{\frac{k}{1+k}} \text{cn}\left(\sqrt{\frac{2}{k}}x,\kappa\right), \quad \kappa=\sqrt{\frac{1-k}{2}}, 
\end{align}
\end{subequations}
and $k \in (0,1)$, $M=2\alpha_2$, $v=4\alpha_3$, $b=2\alpha_2$, $\alpha_2=1$ and $\alpha_3$ is the arbitrary constant.  It follows from (\ref{HE2}) and (\ref{HE3}) that $|r(x,t)|$=$|r(x+L,t)|$=$|r(x,t+T)|$, \;\; $(x,t) \in \mathbb{R}^2$, with the fundamental periods $\displaystyle L=\frac{4K(\kappa)}{\sqrt{1+k}}$ and $\displaystyle T=2K(k)$ for (\ref{HE2}) and $L=2\sqrt{2}K(\kappa)$ and $T=4K(k)$ for (\ref{HE3}), where $K(k)$ represents the complete elliptic integral of the first kind with the elliptic modulus parameter $\kappa$.  In Fig. \ref{dnnfig2}, we display the qualitative profiles of double-periodic wave coming out from the solutions (\ref{HE2}) and (\ref{HE3}) for the modulus $k=0.9$ and $\alpha_2=1$ with two different values of $\alpha_3$.  In Figs. \ref{dnnfig2}(a)-\ref{dnnfig2}(b), double-periodic waves of (\ref{HE2}) are shown for $\alpha_3=1/6$ (or) $0.1667$ and $\alpha_3=0.85$,  respectively.  Figures \ref{dnnfig2}(c) and \ref{dnnfig2}(d) represent the double-periodic profiles of (\ref{HE3}) for $\alpha_3=0.1667$ and $\alpha_3=0.5$, respectively.  From these plots, we observe that double-periodic wave is more localized for low values of $\alpha_3$ when compared to the higher values of $\alpha_3$ which can be visualized from Figs. \ref{dnnfig2}(a) and \ref{dnnfig2}(c).  We also notice that the orientation of double-periodic wave changes when we vary the value of the arbitrary parameter $\alpha_3$.

We intend to construct RWs on the double periodic profiles given in (\ref{HE2}) and (\ref{HE3}).  However, these solutions are travelling wave solutions.  We invoke the method of nonlinearization of Lax pairs \cite{zhou,zhou1,chen1,chen3} in order to identify the eigenvalues for which these solutions appear.  In this procedure one usually identifies the Hamiltonian associated with the spatial and temporal parts of the Lax pair by introducing a constraint between the potential $r(x,t)$ and the eigenfunctions. In our case, the constraint reads $r = \langle \textbf{f}, \textbf{f} \rangle: = f_{1}^{2}+{\bar{g}_{1}}^{2}+f_{2}^{2}+{\bar{g}_{2}}^{2}$,  where the functions $(f_1, g_1, f_2, g_2)$ are nothing but the solutions of the Lax pair Eqs. (\ref{HE4}) and (\ref{HE5}) and bar denotes the complex conjugate of the respective function.

After introducing this constraint we can proceed to construct a set of ordinary differential equations (ODEs) with $r(x,t)$ as the dependent variable.  The order and number of these ODEs normally extends depending upon the number of unknowns to be determined.  In our case, we have four unknowns, namely $(f_1, g_1, f_2, g_2)$ and so we sequentially go upto fourth order ODE. The last differential equation turns out to be the fourth-order Lax-Novikov equation in the hierarchy of stationary NLS equation.  The fourth-order complex-valued Lax-Novikov equation is integrable with two complex-valued constants of motion \cite{chen4}. From these integrals of motion, we identify a set of admissible eigenvalues $\lambda_1$ and $\lambda_2$ which are symmetric about the imaginary axis.  Once the eigenvalues are fixed, substituting the double periodic solution and the obtained squared eigenfunctions in the one-fold Darboux transformation formula (\ref{HE6}), we can create double periodic solution as the background for the considered equation.  The resulting solution is periodic both in space and time. We then proceed to construct RWs on top of this double periodic waves.  To  achieve this we consider a second linearly independent solution for the Lax system (\ref{HE4}) and (\ref{HE5}) with the same set of eigenvalues and eigenfunctions with an unknown function. We determine this unknown function by substituting the considered second independent solution back in the Lax system and integrating the underlying equations. Substituting back the second solution in the one-fold Darboux transformation formula (\ref{HE6}), we obtain the RW solution on top of the double-periodic wave background.        
\section{Demonstration on Hirota equation}
In this section, we demonstrate the method of constructing RW solutions on the double periodic background for the Hirota equation.   In Ref. \cite{chen4}, the authors have constructed the rogue wave solutions on the double periodic background for the focusing NLS equation.  We adopt the same procedure and derive the RW solution on the double periodic background to Eq. (\ref{HE1}). While implementing this procedure, we observe that Hirota equation also yields the same results upto section III in Ref. \cite{chen4} with $\alpha_2=1$.  While performing the procedure of nonlinearization of Lax pair for the Eqs. (\ref{HE4}) and (\ref{HE5}) with the constraint $r = f_{1}^{2}+{\bar{g}_{1}}^{2}+f_{2}^{2}+{\bar{g}_{2}}^{2}$ in the time part, we come across certain differences.  For instance, substituting the assumed constraint into Eqs. (\ref{HE4}) and (\ref{HE5}), we obtain the following finite-dimensional Hamiltonians, namely
\begin{equation}
\frac{df_j}{dx}=\frac{\partial H_0}{\partial g_j},\qquad \frac{dg_j}{dx}=-\frac{\partial H_0}{\partial f_j},\qquad j=1,2,
\label{HE9}
\end{equation}
where
\begin{equation}
H_0=\langle \Lambda \textbf{f,g}\rangle+\frac{1}{2}\langle \textbf{f,f}\rangle \langle \textbf{g,g}\rangle,
\label{HE10}
\end{equation}
and 
\begin{eqnarray}
\frac{df_j}{dt}=\frac{\partial H_1}{\partial g_j},\qquad \frac{dg_j}{dt}=-\frac{\partial H_1}{\partial f_j},\qquad j=1,2.
\label{HE11}
\end{eqnarray}
where
\begin{align}
\label{HE12}
H_1=&\frac{1}{4}[i\langle \textbf{f,g}\rangle+\frac{2}{3}{\langle \textbf{f,f}\rangle}^2{\langle \textbf{g,g}\rangle}^2+i\frac{1}{2} \langle \textbf{f,f}\rangle \langle \Lambda \textbf{g,g}\rangle  \nonumber \\&+i\frac{1}{2} \langle\Lambda \textbf{f,f}\rangle \langle  \textbf{g,g}\rangle  +\frac{4}{3} \langle\Lambda \textbf{f,g}\rangle \langle  \textbf{f,f}\rangle \langle  \textbf{g,g}\rangle +2 i\langle \Lambda^2 \textbf{f,g}\rangle\nonumber \\&+\frac{2}{3}\langle \Lambda^2 \textbf{g,g}\rangle\langle \textbf{f,f}\rangle+\frac{2}{3}\langle \Lambda^2 \textbf{f,f}\rangle\langle \textbf{g,g}\rangle  +\frac{8}{3}\langle \Lambda^3 \textbf{f,g}\rangle  \nonumber \\&-\frac{i}{3}\langle \textbf{f,f}\rangle\langle \textbf{f},\textbf{g}^3\rangle -\frac{i}{3}\langle \textbf{g,g}\rangle\langle \textbf{f}^3,\textbf{g}\rangle-\frac{1}{6}{\langle \textbf{f,f}\rangle}^{2}\langle \textbf{g}^2,\textbf{g}^2\rangle  \nonumber \\&-\frac{1}{6}\langle\textbf{g,g}\rangle\langle \textbf{f}^2,\textbf{f}^2\rangle -\frac{4}{9}\langle \textbf{f,f}\rangle\langle \Lambda \textbf{f},\textbf{g}^3\rangle-\frac{4}{9}\langle \textbf{g,g}\rangle\langle \Lambda \textbf{f}^3,\textbf{g}\rangle]. 
\end{align}
In Eqs. (\ref{HE10}) and (\ref{HE12}), the notations \textbf{f}, \textbf{g} and \textbf{$\Lambda$} denote \textbf{f}=$(f_1,f_2,\bar{g}_1,\bar{g}_2)^T$, \textbf{g}=$(g_1,g_2,-\bar{f}_1,-\bar{f}_2)^T$ and \\ $\Lambda=\text{diag}(\lambda_1,\lambda_2,-\bar{\lambda}_1,-\bar{\lambda}_2)$, respectively.  One may notice that the Hamiltonian (\ref{HE12}) which comes from the time evolution part of Lax pair of the Hirota equation differs from the NLS equation \cite{chen4}.

\begin{figure*}[!ht]
	\begin{center}
		\subfigure[]
		{
		\resizebox{0.4\textwidth}{!}{\includegraphics{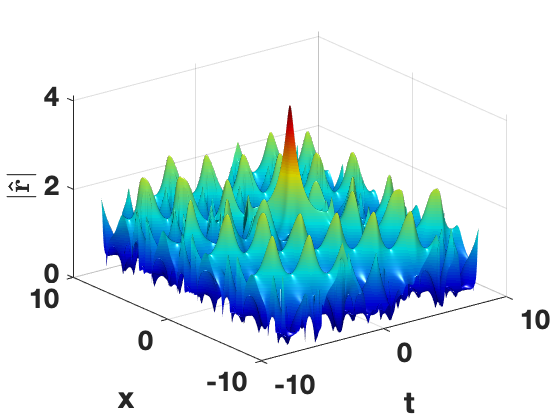}}
		\label{fig:21}
	}~~ 
\subfigure[]
{
		\resizebox{0.4\textwidth}{!}{\includegraphics{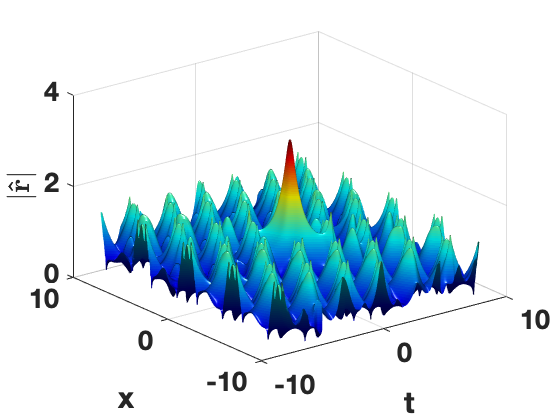}} 
		\label{fig:22}
	}\\
\subfigure[]
{
		\resizebox{0.4\textwidth}{!}{\includegraphics{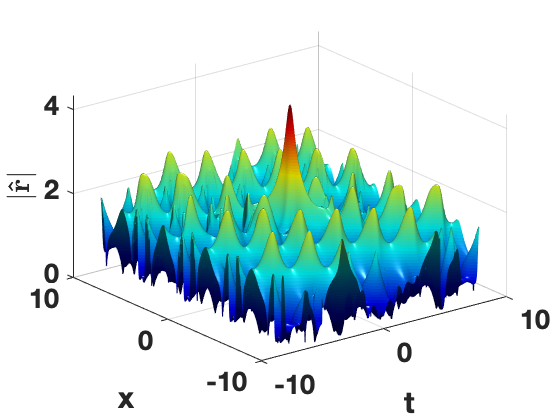}}
		\label{fig:23}
	}
	\end{center}
	\vspace{-0.2cm}
	\caption{RWs on the background of the double-periodic solution (\ref{HE2}) for the Hirota Eq. (\ref{HE1}) with $k=0.8$ and $\alpha_3=0.1667$ for three different eigenvalues (a) $\lambda_1=\sqrt{\tau_1}$, (b) $\lambda_2=\sqrt{\tau_2}$ and (c) $\lambda_3=\sqrt{\tau_3}$.}
	\label{dnnfig3}
\end{figure*}

\begin{figure*}[!ht]
	\begin{center}
		\subfigure[]
		{
		\resizebox{0.4\textwidth}{!}{\includegraphics{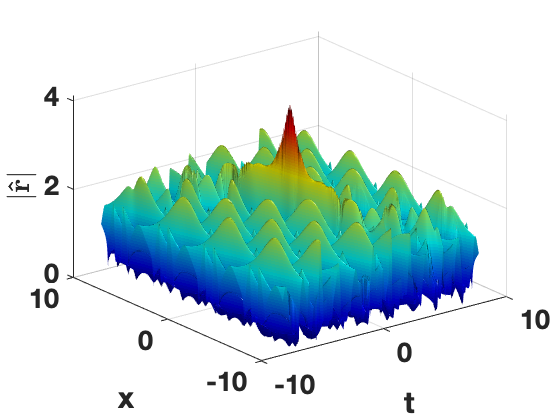}}
		\label{fig:21}
	}~~ 
\subfigure[]
{
		\resizebox{0.4\textwidth}{!}{\includegraphics{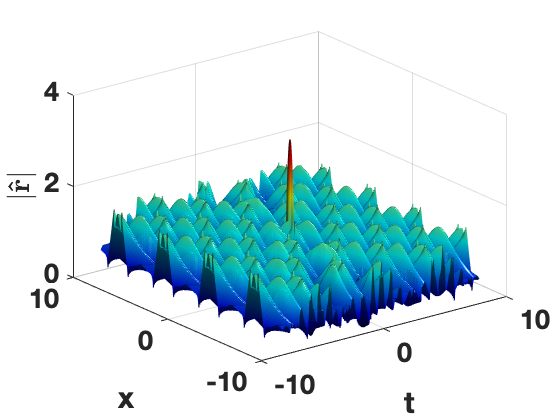}} 
		\label{fig:22}
	}\\
\subfigure[]
{
		\resizebox{0.4\textwidth}{!}{\includegraphics{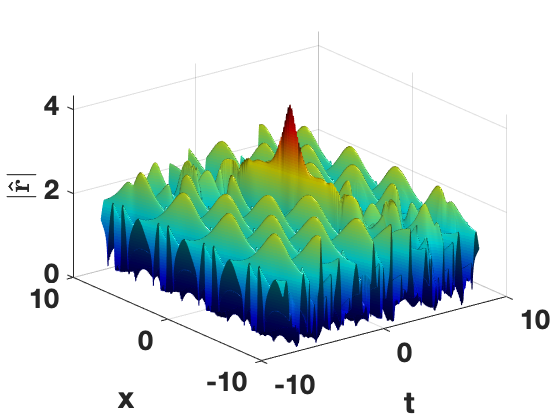}}
		\label{fig:23}
	}
	\end{center}
	\vspace{-0.2cm}
	\caption{RWs on the background of the double-periodic solution (\ref{HE2}) for the Hirota Eq. (\ref{HE1}) with $k=0.8$ and $\alpha_3=0.85$  for three different eigenvalues (a) $\lambda_1=\sqrt{\tau_1}$, (b) $\lambda_2=\sqrt{\tau_2}$ and (c) $\lambda_3=\sqrt{\tau_3}$.}
	\label{dnnfig3ab}
\end{figure*}

\begin{figure*}[!ht]
	\begin{center}
			\subfigure[]
			{
		\resizebox{0.4\textwidth}{!}{\includegraphics{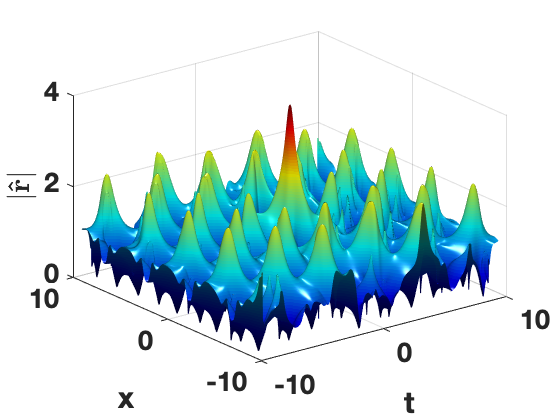}}
		\label{fig:31}
	}~~ 
	\subfigure[]
	{
		\resizebox{0.4\textwidth}{!}{\includegraphics{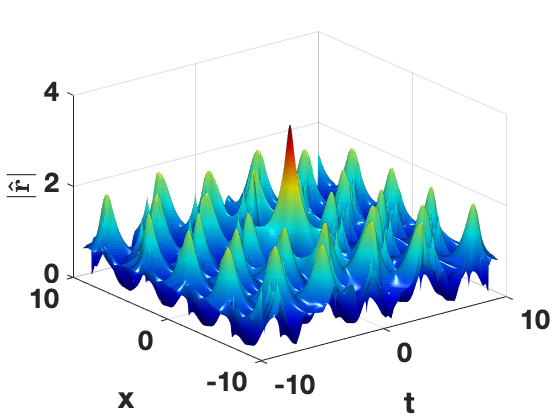}} 
		\label{fig:32}
	}\\
	\subfigure[]
	{
		\resizebox{0.4\textwidth}{!}{\includegraphics{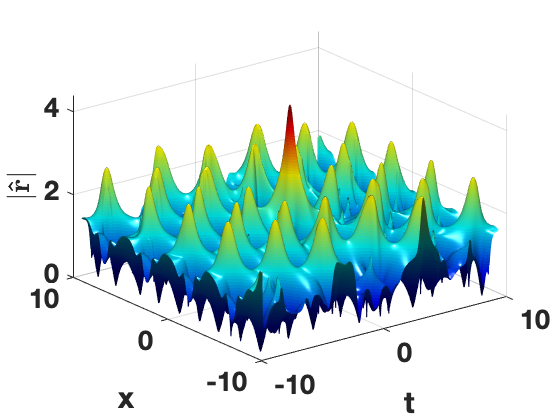}}
		\label{fig:33}
	}
	\end{center}
	\vspace{-0.2cm}
	\caption{RWs on the background of the double-periodic solution (\ref{HE2}) for the Hirota Eq. (\ref{HE1}) with $k=0.95$ and $\alpha_3=0.1667$ and for eigenvalues (a) $\lambda_1=\sqrt{\tau_1}$, (b) $\lambda_2=\sqrt{\tau_2}$ and (c) $\lambda_3=\sqrt{\tau_3}$.}
	\label{dnnfig3a}
\end{figure*}

The squared eigenfunctions $f_1^2$ and $g_1^2$ which appear in the  Eq. (\ref{HE6}) can be obtained by substituting the expressions (\ref{HE2}) and (\ref{HE3}) and their derivatives in the following equations:  
\begin{align}
\label{HE66}
f_1^2= & \frac{\lambda_1}{4(\lambda_1+\bar{\lambda}_1)(\lambda_1-\lambda_2)(\lambda_1+\bar{\lambda}_2)}\nonumber\\&~~~\times\left[r''+2|r|^2 r+4(b+\lambda_1^2)r+2\lambda_1 r'\right], \notag \\
g_1^2= & \frac{\lambda_1}{4(\lambda_1+\bar{\lambda}_1)(\lambda_1-\lambda_2)(\lambda_1+\bar{\lambda}_2)}\nonumber\\&~~~\times\left[\bar{r}''+2|r|^2\bar{r}+4(b+\lambda_1^2)\bar{r}-2\lambda_1 \bar{r}'\right], \notag \\
f_1g_1= & -\frac{\lambda_1}{4(\lambda_1+\bar{\lambda}_1)(\lambda_1-\lambda_2)(\lambda_1+\bar{\lambda}_2)}\nonumber\\&~~~\times\left[r'\bar{r}-r\bar{r}'+2\lambda_1 (2b+2\lambda_1^2+|r|^2)\right].
\end{align}
(since the details of getting (\ref{HE66}) have already been given in Ref. \cite{chen4} we do not reproduce them here).

Similarly we can determine the expressions of squared eigenfunctions, namely $f_2^2$, $g_2^2$ and $f_2g_2$ from the above equations by replacing $\lambda_1$ by $\lambda_2$ and $\lambda_2$ by $\lambda_1$.  We consider two different sets of three admissible pairs of eigenvalues among $\pm \lambda_1$, $\pm \lambda_2$ and $\pm \lambda_3$ which were found in Ref. \cite{chen4} for the family of double-periodic wave solutions given in Eqs. (\ref{HE2}) and (\ref{HE3}).  Each pair of eigenvalues can be taken in place of $\lambda_1$ and $\lambda_2$.  In the first set, all the eigenvalues are considered to be real, that is $\lambda_1=\pm\sqrt{\tau_1}$, $\lambda_2=\pm\sqrt{\tau_2}$, and $\lambda_3 =\pm\sqrt{\tau_3}$ where $\tau_1$ varies from $0$ to $1$, $\tau_2=\tau_3-\tau_1$ and $\tau_3=1$.  In the second set, it is considered that one of the eigenvalues $\lambda_1=\pm\sqrt{\tau_1}$ is positive and the other two eigenvalues are complex-conjugate, namely $\lambda_2=\pm \sqrt{\beta + i \gamma}$ and $\lambda_3=\pm \sqrt{\beta - i \gamma}$, where $\tau_1=2\beta$ and $\beta$ varies from $0$ to $0.5$.   Now, substituting the resultant expressions coming out from (\ref{HE66}) and considering three admissible pairs of eigenvalues given above in (\ref{HE6}), we can set up the double-periodic wave pattern as background for the Hirota Eq. (\ref{HE1}).

\begin{figure*}[!ht]
	\begin{center}
			\subfigure[]
			{
		\resizebox{0.4\textwidth}{!}{\includegraphics{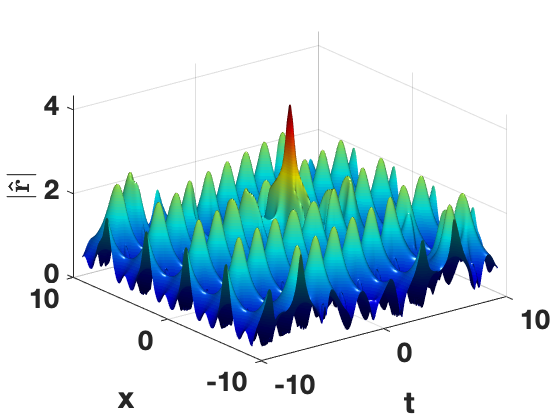}}
		\label{fig:41}
	}~~ 
	\subfigure[]
	{
		\resizebox{0.4\textwidth}{!}{\includegraphics{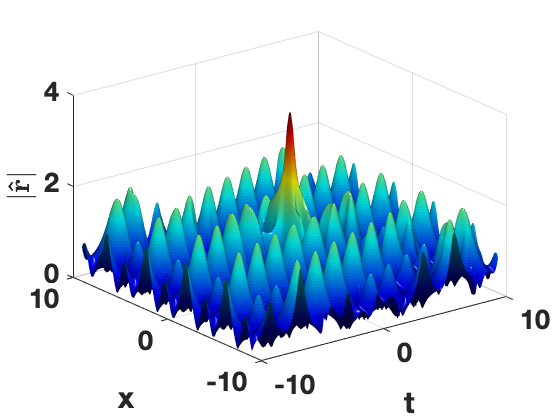}}
		\label{fig:42}
	}
	\end{center}
	\vspace{-0.2cm}
	\caption{RWs on the background of the double-periodic solution (\ref{HE3}) for the Hirota Eq. (\ref{HE1}) with $k=0.9$ and $\alpha_3=0.1667$ for complex conjugate eigenvalues (a) $\lambda_1=\sqrt{\tau_1}$ and (b) $\lambda_2=\sqrt{\beta + i \gamma} $.}
	\label{dnnfig4}
\end{figure*}

\begin{figure*}[!ht]
	\begin{center}
			\subfigure[]
			{
		\resizebox{0.4\textwidth}{!}{\includegraphics{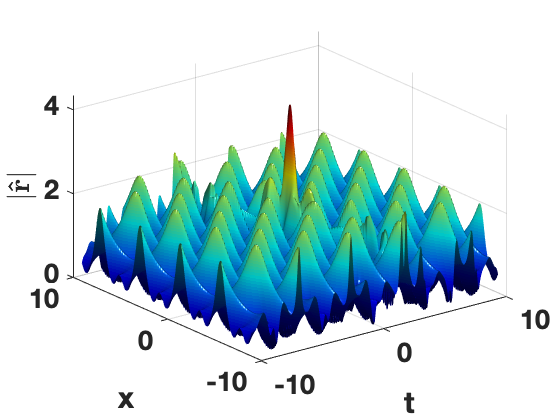}}
		\label{fig:41}
	}~~ 
	\subfigure[]
	{
		\resizebox{0.4\textwidth}{!}{\includegraphics{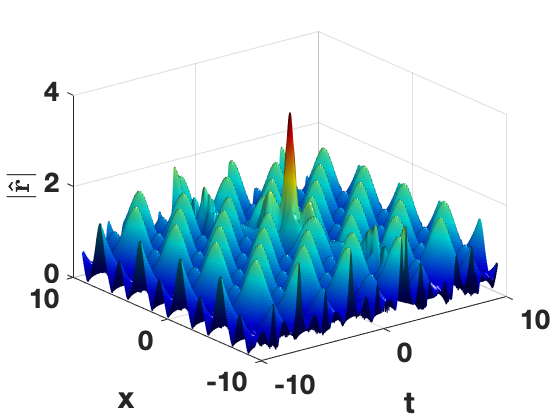}}
		\label{fig:42}
	}
	\end{center}
	\vspace{-0.2cm}
	\caption{RWs on the background of the double-periodic solution (\ref{HE3}) for the Hirota Eq. (\ref{HE1}) with $k=0.9$ and $\alpha_3=0.5$ for complex conjugate eigenvalues (a) $\lambda_1=\sqrt{\tau_1}$ and (b) $\lambda_2=\sqrt{\beta + i \gamma} $.}
	\label{dnnfig4a}
\end{figure*}

To generate RWs on this double periodic background, we construct a second linearly independent solution for the Lax pair Eqs. (\ref{HE4}) and (\ref{HE5}) with the same eigenvalues. The second linearly independent solution which we intend to construct should be non-periodic and grow linearly in $x$ and $t$.  We consider the second linearly independent solution $\varphi=(\hat{f}_1,\hat{g}_1)^T$ for the Lax pair Eqs. (\ref{HE4}) and (\ref{HE5}) in the form 
\begin{align}
\label{HE67}
\hat{f}_1 = f_1 \eta_1-\frac{2\bar{g}_1}{|f_1|^2+|g_1|^2}, \quad \hat{g}_1 =  g_1 \eta_1+\frac{2\bar{f}_1}{|f_1|^2+|g_1|^2},
\end{align}
where $\eta_1$ is the unknown function to be determined.  The Wronskian between the two solutions (\ref{HE6}) and (\ref{HE67}) is nonzero, 
\begin{align}
\label{HE67a}
f_1\hat{g}_1-\hat{f}_1g_1=&f_1\left(g_1 \eta_1+\frac{2\bar{f}_1}{|f_1|^2+|g_1|^2}\right)\nonumber\\&-g_1\left(f_1 \eta_1-\frac{2\bar{g}_1}{|f_1|^2+|g_1|^2}\right)=2,
\end{align}
which in turn proves that the second solution is linearly independent from the first one.  

\begin{figure*}[!ht]
	\begin{center}
		\subfigure[]
		{
		\resizebox{0.4\textwidth}{!}{\includegraphics{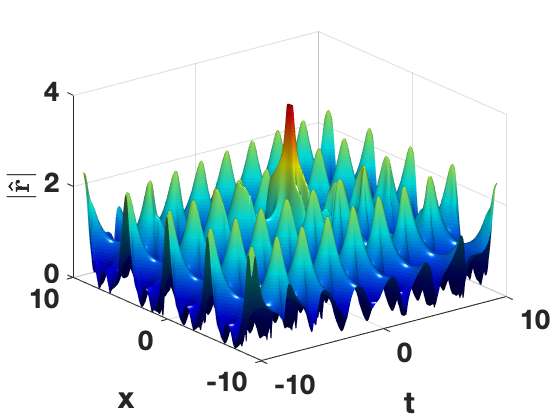}}
		\label{fig:51}
	}~~ 
\subfigure[]
{
		\resizebox{0.4\textwidth}{!}{\includegraphics{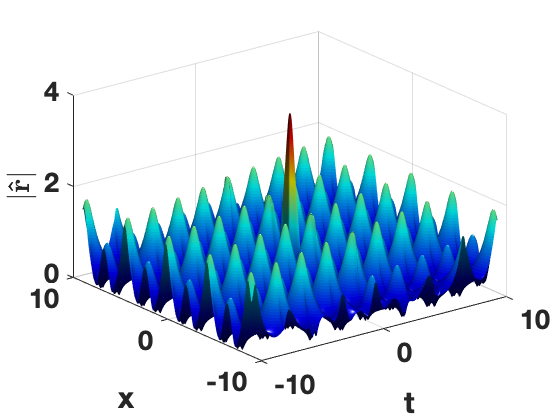}}
		\label{fig:52}
	}
	\end{center}
	\vspace{-0.2cm}
	\caption{RWs on the background of the double-periodic solution (\ref{HE3}) for the Hirota Eq. (\ref{HE1}) with $k=0.95$ and $\alpha_3=0.1667$ for complex conjugate eigenvalues (a) $\lambda_1=\sqrt{\tau_1}$ and (b) $\lambda_2=\sqrt{\beta + i \gamma} $.}
	\label{dnnfig5}
\end{figure*}

Substituting (\ref{HE67}) into (\ref{HE4}) and utilizing (\ref{HE2}) for $\varphi =(\hat{f}_1,\hat{g}_1)^T$, we obtain the following first-order partial differential equation for the unknown function $\eta_1$, that is 
\begin{align}
\label{HE68}
\frac{\partial \eta_1}{\partial x} = W_1:= & -\frac{4(\lambda_1+\bar{\lambda}_1)\bar{f}_1\bar{g}_1}{\left(|\hat{f}_1|^2+|\hat{g}_1|^2\right)^2}.
\end{align}
\par Similarly, inserting (\ref{HE67}) into (\ref{HE5}) and utilizing (\ref{HE5}) for $\varphi =(\hat{f}_1,\hat{g}_1)^T$, we obtain another first-order partial differential equation for $\eta_1$ in time derivative, namely
\begin{align}
\label{HE68a}
\frac{\partial \eta_1}{\partial t} = W_2: =&  \frac{2 (\lambda_1+\bar{\lambda}_1)S_1\bar{f}_1^2}{(|f_1|^2+|g_1|^2)^2}
+\frac{2 (\lambda_1+\bar{\lambda}_1)S_2\bar{g}_1^2}{(|f_1|^2+|g_1|^2)^2}\notag \\ &
-\frac{4 (\lambda_1+\bar{\lambda}_1)S_3\bar{f}_1\bar{g}_1}{(|f_1|^2+|g_1|^2)^2},
\end{align}
where $S_1=2 r_x\alpha_3+(i+4\alpha_3(\lambda_1- \bar{\lambda}_1))r$, $S_2=-2\bar{r}_x\alpha_3+(i+4\alpha_3(\lambda_1- \bar{\lambda}_1))\bar{r}$ and $S_3=2|r|^2\alpha_3+4\alpha_3\lambda_1^2+(i-4\alpha_3\bar{\lambda}_1)(\lambda_1-\bar{\lambda}_1)$.  
The first-order equations (\ref{HE68}) and (\ref{HE68a}) are compatible with each other ($W_{1t}=W_{2x}$).  Upon integrating the expressions (\ref{HE68}) and (\ref{HE68a}), we obtain  
\begin{align}
\label{HE69}
\eta_1(x,t)= & \int_{x_0}^{x} W_1(x',t)dx'+ \int_{t_0}^t W_2(x_0,t')dt',
\end{align}
where $(x_0,t_0)$ is arbitrarily fixed. It is very difficult to integrate Eq. (\ref{HE69}) and obtain an explicit expression for $\eta_1(x,t)$ analytically.  So we evaluate the expression (\ref{HE69}) numerically by Newton-Raphson method.

Substituting the double periodic solutions $r(x,t)$ (given in (\ref{HE2}) and (\ref{HE3})) and the eigenfunctions of the second solution $\varphi =(\hat{f}_1,\hat{g}_1)^T$ of the linear equations (\ref{HE4})-(\ref{HE5}) with $\lambda=\lambda_1$ in the one-fold Darboux transformation formula (\ref{HE6}), we obtain a new form of solution (RW solution on the double periodic background) to the Hirota Eq. (\ref{HE1}) in the form 
\begin{align}
\label{HE70}
\hat{r}(x,t)= & r(x,t)+\frac{2(\lambda_1+\bar{\lambda}_1)\hat{f}_1\bar{\hat{g}}_1}{|\hat{f}_1|^2+|\hat{g}_1|^2},  
\end{align}
where $\hat{f}_1$ and $\hat{g}_1$ are given in (\ref{HE67}). We note here that this second solution differs from the one reported for the NLS equation.

In Figs. \ref{dnnfig3}(a)-(c), we show the surface plots of $|\hat{r}|$ of RWs on the double periodic background using the solution ({\ref{HE2}}) with $k=0.8$ and $\alpha_3=1/6$ (or) $0.1667$ for three real eigenvalues, namely $\lambda_1=\sqrt{\tau_1}$, $\lambda_2=\sqrt{\tau_2}$ and $\lambda_3=\sqrt{\tau_3}$ with $\tau_1=0.8$, $\tau_3=1$ and $\tau_2=\tau_3-\tau_1$.  In this figure, in all the cases, RW attains their maximum amplitude at their origin, that is $(x_0, t_0)=(0,0)$.  The amplitude of RWs is different in each case and is found to be $|\hat{r}| \approx 4.13, 3.236$ and $4.342$ for the eigenvalues $\lambda_1$, $\lambda_2$ and $\lambda_3$, respectively.  The outcome is shown in Figs. \ref{dnnfig3}(a), \ref{dnnfig3}(b) and \ref{dnnfig3}(c).  When we increase the value of the arbitrary parameter $\alpha_3$ to $0.85$, the amplitude of RWs remains the same as in the case of $\alpha_3=0.1667$.  But we observe that the double-periodic background waves are localized in different positions in the $(x-t)$ plane which can be visualized from the Figs. \ref{dnnfig3ab}(a)-(c).  For $k=0.95$ and $\alpha_3=0.1667$, the surface plots of $|\hat{r}|$ of RWs on the double periodic background using the solution ({\ref{HE2}}) with three real eigenvalues, namely $\lambda_1=\sqrt{\tau_1}$, $\lambda_2=\sqrt{\tau_2}$ and $\lambda_3=\sqrt{\tau_3}$ with $\tau_1=0.655$, $\tau_3=1$ and $\tau_2=\tau_3-\tau_1$ are displayed in Figs. \ref{dnnfig3a}(a)-(c).  For this set of eigenvalues also, we observe that the amplitude of RWs exhibits a maximum value at their origins.  The amplitudes of RWs for the eigenvalues $\lambda_1$, $\lambda_2$ and $\lambda_3$ are $|\hat{r}| \approx 4.015, 3.571$ and $4.397$ which are shown in Figs. \ref{dnnfig3a}(a), \ref{dnnfig3a}(b) and \ref{dnnfig3a}(c), respectively.  From these observations, we infer that the amplitude of RWs increases in the order of eigenvalues as $\lambda_2 < \lambda_1 < \lambda_3$ in the $x-t$ plane.   The amplitude of RWs for the eigenvalue $\lambda_3$ is higher than that of the other two eigenvalues ($\lambda_1$ and $\lambda_2$).

Figure \ref{dnnfig4} represents the surface plots of $|\hat{r}|$ of RWs on the double periodic background using the solution ({\ref{HE3}}) with $k=0.9$ for (a) $\lambda_1=\sqrt{\tau_1}$, $\tau_1=2\beta$ and (b) $\lambda_2=\sqrt{\beta + i \gamma} $, $\beta=0.45$ and $\gamma=0.217945$. From Figs. \ref{dnnfig4}(a) and \ref{dnnfig4}(b), one can visualize the amplitude of RWs attains the maximum value $|\hat{r}| \approx 4.35$ and $3.841$ for the real eigenvalue $\lambda_1$ and the complex eigenvalue $\lambda_2$, respectively.  The surface plots of $|\hat{r}|$ for the same set of eigenvalues but for the choice of the arbitrary parameter, $\alpha_3=0.5$ are presented in Figs. \ref{dnnfig4a}(a) and \ref{dnnfig4a}(b).  Here we notice that the localization of double-periodic background waves occurs in different orientations whereas the amplitudes of RWs remain the same as in the case of $\alpha_3=0.1667$. The RWs on the double periodic background which are shown in Figs. \ref{dnnfig5}(a)-(b) are drawn for the same set of values as given in Fig. \ref{dnnfig4} with $k=0.95$ and $\alpha_3=0.1667$.  The maximum amplitudes of RWs for the eigenvalues $\lambda_1$ and $\lambda_2$ are $|\hat{r}| \approx 4.025$ and $3.811$.  This is shown in Figs. \ref{dnnfig5}(a) and \ref{dnnfig5}(b).  We  conclude that the amplitude of RWs for the case of real eigenvalue ($\lambda_1$) is higher than that of the complex eigenvalue ($\lambda_2$).  
\section{Summary}
In this work, we have constructed RW solutions on the double-periodic background for the Hirota equation in an algebraic way.  One of the efficient tools to derive RW solution for the given nonlinear partial differential equation is the Darboux transformation method.  Through this method, one can construct the desired solution by appropriately choosing the seed solution.  However, it is very difficult to integrate the underlying equations arising from Lax pair when we consider double periodic solution as seed solution.  To circumvent this difficulty, we considered another procedure through which the necessary expressions for the squared eigenfunctions and the eigenvalues that appear in the one fold Darboux transformation formula can be identified.  The procedure which we brought-in to identify the eigenfunctions and eigenvalues is the method of nonlinearization of Lax pairs introduced long ago in a different context in the nonlinear dynamics literature.  With the help of this method and the spectral theory of Lax pairs, we have captured the eigenvalues and the squared eigenfunctions which appear in the one-fold Darboux transformation formula.  With the obtained expressions, we have created the double periodic waves as background.  In the second step, we have created the RW on top of this background.  We have presented the localized structures for two different values of the arbitrary parameter each one with two sets of eigenvalues.  Our studies reveal that the amplitudes of RWs increase in the order of eigenvalues as $\lambda_2 < \lambda_1 < \lambda_3$ in the $(x-t)$ plane.  Our results also show that when we vary the value of the arbitrary parameter $\alpha_3$ the amplitudes of RWs remain the same and localization of the double-periodic background waves occur in different orientations in the $(x-t)$ plane. The application of this approach to few other evolutionary equations is under progress.

\section*{Acknowledgments}
NS thanks the University for providing University Research Fellowship.  KM wishes to thank the Council of Scientific and Industrial Research, Government of India, for providing the Research Associateship under the Grant No. 03/1397/17/EMR-II. The work of MS forms part of a research project sponsored by National Board for Higher Mathematics, Government of India, under the Grant No. 02011/20/2018NBHM(R.P)/R\&D 24II/15064.  The authors thank Prof. Dmitry E. Pelinovsky for helping to evaluate the numerical part of this work.

\end{document}